\documentclass[12pt,preprint]{revtex4}

\usepackage{graphicx}

\usepackage{array}

\begin{document}

\title{Relation between the chemical force and the tunneling current in atomic point contacts: a~simple model}
\author{Pavel Jel\'{i}nek}
\email{jelinekp@fzu.cz}

\author{Martin Ondr\'{a}\v{c}ek}
\affiliation{Institute of Physics, Academy of Sciences of the Czech
Republic, Cukrovarnick\'{a} 10, 1862 53, Prague, Czech Republic}

\author{Fernando Flores}
\affiliation{Departamento de F\'{\i}sica Te\'orica de la Materia Condensada,
Universidad Aut\'onoma de Madrid, E-28049 Spain}
%\begin{abstract}

%\end{abstract}

\maketitle

\today

\section{Abstract}
The relation between the tunneling current $I_t$ and the interaction energy/force $E$ in an atomic-scale contact is discussed in a frame of a~theoretical model established here. 
According to our model, we predict an~existence of two characteristic scaling regimes, where the tunneling current is either proportional to the interaction energy $I_t \sim E$ 
or to the the square of the interaction energy, $I_t \sim E^2$.  We show that the existence of a~given regime is basically controlled by two parameters: 
(i) the energy degeneracy $\Delta$ and (ii) the hopping $t$ between electronic levels involved in the interaction process.  In addition, we discuss the validity of the Bardeen method 
to determine the tunneling current in the short tip-sample distances.

\section{Introduction}

The characterization and modification of surfaces and nanostructures is one of the most pressing challenges of Nanoscience these days. 
Scanning Probe Microscopy (SPM) has already achieved outstanding performance in this field. In particular, Scanning Tunneling Microscopy (STM) \cite{STM} 
and Atomic Force Microscopy (AFM) \cite{AFM} are two leading experimental tools that have been used for this purposes. The first, sensing the tunneling current between 
a~probe  and a sample, has been already converted to a standard technique widely used to image, characterize and modify objects at the atomic scale. 
However, universal application of this method is limited by the requirement of a~conductive sample to detect the tunneling current. This serious drawback is not an issue anymore for AFM, which senses the force acting between the tip and the sample.
This is one of the reasons why the latter method has been widely adopted in recent years, when the AFM method became widespread in different scientific fields (e.g. biology, chemistry and physics). 
%Sensing the force, AFM has been converted into a tool capable to measure forces at atomic scale. 

Further proliferation of SPM methods is closely tied to a~detailed understanding of undergoing processes during imaging or manipulation. In particular,
the tip-sample interaction is a key for a better control and interpretation of measurements. Introduction of a~so-called qPlus sensor \cite{qPLus} allowed simultaneous detection of both tunneling current and forces.
This approach opens up an advanced way to characterize objects at the atomic scale combining different detection signals at the same time. 
Moreover, it provides a direct and precise way to study the relation between the chemical force and the tunneling current.  This question has already received 
a~lot of attention from both theory \cite{Hofer03,Chen05,Ternes11} and experiment \cite{Loppacher00, Hembacher05, Bollinger04,Schirmeisen00, Sun05, Konig09}. 
Recently, simultaneous measurements of both atomic force and tunneling current with high precision have been reported \cite{Sawada09a, Ternes08, Ternes11}.
Therefore it is evident we have come to the stage at which precise simultaneous measurements of the force and the current become the standard.
This situation calls for deeper theoretical understanding of how the tunneling current and force behave in different interaction regimes.
In particular, the question how both the tunneling current and the chemical force are correlated with each other has received certain attention and it is still not fully resolved \cite{Hofer03, Chen05}.

% another applications
Simultaneous acquisition of the electron current and the force between nanoscale objects would elaborate our knowledge of their physical and material properties 
and undergoing processes in atomic scale.  Indeed, direct comparison of these two quantities measured on single gold atom wires during a breaking process has already
shed more light on the origin of conductance quantization in atomic nanowires~\cite{Bollinger01} and the relation between their mechanical and transport properties.
Application of the technique would also bring valuable information about the binding mechanism and electron transport through a~single molecule anchored to metallic electrodes \cite{Cui01,Nitzan03,VenkataramanNature06}. Very recently this kind of experiment has been already realized \cite{Tautz_PRB11}.  
Devices integrating electrical and mechanical functionality on the nanoscale, so called NEMS (NanoElectroMechanical Systems) systems are another research area 
(see e.g.~\cite{Schwab05}) where the simultaneous measurement of the current and force could open new perspectives. 

Therefore, to understand the the correlation between force and current in atomic scale, we provide here a~simple model. In this model (i) we discus the implication of 
the strong interaction regime on the electron current through the tip-sample junction and (ii) we explain the relation between the chemical force and the tunneling current 
in the weakly interacting regime. In particular, we will show in this paper that the relation between the tunneling current and the chemical force can be used as an~indicator 
of the quantum degeneracy between the frontier electronic states of the tip and the sample involved in the electron transfer.

\section{Model}
The aim of this paper is to get more insight into the relation between the tunneling current and the chemical force 
arising during the formation of an~atomic contact between an SPM probe and a~surface atom. We will introduce a simple model, describing both 
the electron tunneling and the interaction energy between two bodies representing the SPM tip and the sample.
We will denote the two quantum-mechanical systems corresponding to the two interacting bodies as $\alpha, \beta$.
The systems are described by eigenfunctions $\phi_{\alpha}$, $\phi_{\beta}$ and eigenenergies $\epsilon_{\alpha}$, $\epsilon_{\beta}$
satisfying the Schr\"{o}dinger equation for the respective decoupled system $\alpha$ or $\beta$.
Next, we will consider only the interaction between the two outermost atoms on the tip and the sample. Indeed, this approximation can be justified by the exponential decay of wave-functions outside the surface, which limits the interaction 
between the two bodies to only the outermost atomic layers. In particular, recently reported single-atom chemical identification is based on strong locality of the chemical force,
almost completely driven by the apex atom of the tip and the nearest surface atom \cite{Sugimoto-Nature07,Pou-Nanotechnology09}. The same reasoning 
is also plausible for the tunneling current \cite{Chen93}. 

%\begin{figure}
%\includegraphics[width=\columnwidth]{Figs/scheme/scheme.pdf}
%\caption{}\label{Fig:scheme}
%\end{figure}

We thus start with two decoupled systems, $\alpha$ (surface) and $\beta$ (tip), which can be described
as follows:
\begin{equation}
\left(-\frac{\hbar^{2}}{2m_{e}}\nabla^{2}+V_{\alpha}(\vec{r}-\vec{R}_{\alpha})\right)\phi_{\alpha}=\epsilon_{\alpha}^{o}\phi_{\alpha}
\label{eq:se1}
\end{equation}
and
\begin{equation}
\left(-\frac{\hbar^{2}}{2m_{e}}\nabla^{2}+V_{\beta}(\vec{r}-\vec{R}_{\beta})\right)\phi_{\beta}=\epsilon_{\beta}^{o}\phi_{\beta},
\label{eq:se2}
\end{equation}
where $\epsilon_{\alpha},\epsilon_{\beta}\,\mathrm{and}\,\phi_{\alpha},\phi_{\beta}\,$ are eigenenergies and 
orthonormalized wave functions of independent systems $\alpha$ and $\beta$, respectively.

%%%
\section{Tunneling current }
In the first part of this section we will focus on the far distance regime, where both the tip and the sample are only weakly disturbed by the interacting potential.
In this regime, the electron transport through the gap is driven by the tunneling process.
Afterwards, we will discuss the close distance regime, in which strong coupling between the two systems, tip and sample, occurrs.  In this regime, three 
main factors influence the electron transport: (i) a contraction of the piezo distance between the tip and sample \cite{Blanco_PRB04,Sugimoto_PRB06}; 
(ii) a change of local density of states due to the formation of the chemical bond between the tip and the sample  \cite{Jelinek_PRL08,Neel07};  
and (iii) a multiple scattering effect of electrons in the gap \cite{Blanco_PRB04}.  
In general, there is no strict definition of a tip-sample distance demarcating these two regimes. Nevertheless, the onset of the short-range chemical force
and the deviation of the I-z curve from the exponential form are signatures of the strong interaction regime.   

%%%%%% 
\subsection {Weakly interacting regime}

If the two systems are brought into contact, an overlap between wave-functions
$\phi_{\alpha},\phi_{\beta}$ arises, giving non-zero probability to
transfer an electron from one side to the other and vice versa. Then the Schr\"{o}dinger equation of the coupled system has the following form:
\begin{equation}
\left(-\frac{\hbar^{2}}{2m_{e}}\nabla^{2}+V_{\alpha}(\vec{r}-\vec{R}_{\alpha})+V_{\beta}(\vec{r}-\vec{R}_{\beta})\right)\psi_{i}=\epsilon_{i}\psi_{i}.
\end{equation}
For simplicity, we consider the potential $V$ as a~sum of potentials $V=V_{\alpha}+V_{\beta}$ that satisfy the condition $V_{\alpha}\cdot V_{\beta}=0$.
This condition minimizes an error introduced by the perturbation theory (for detailed discussion see~\cite{Chen93}). 
In the weakly interacting regime, the tunneling current between two electrodes is commonly described by the Bardeen approach \cite{Bardeen60}
within the second order perturbation theory:

\begin{equation}
I_{t}=\frac {4\pi e}{\hbar}\sum_{\alpha,\beta}|T_{\alpha,\beta}^{B}|^{2}\delta(\epsilon_{\alpha}-\epsilon_{\beta}-eV),
\end{equation}
or in an integral form:
\begin{equation}
I_{t}=\frac {4\pi e}{\hbar}\int_{0}^{eV}|T_{\alpha,\beta}^{B}|^{2} \rho_{\alpha}(\epsilon-eV)\rho_{\beta}(\epsilon) d\epsilon,
\end{equation}
where the tunneling matrix $T_{\alpha,\beta}^{B}$ is expressed via the Bardeen surface integral in terms of unperturbed wave
functions of the original systems $\phi_{\alpha},\phi_{\beta}$:
\begin{equation}
T_{\alpha,\beta}^{B}=\int_{\Omega}(\phi_{\beta}^{*}\nabla\phi_{\alpha}-\phi_{\alpha}\nabla\phi_{\beta}^{*})d\vec{s}\ .
\end{equation}

%%%%%%
\subsection {Strongly interacting regime}
\label{sec:SIR}

\begin{figure}
\includegraphics[width=\columnwidth]{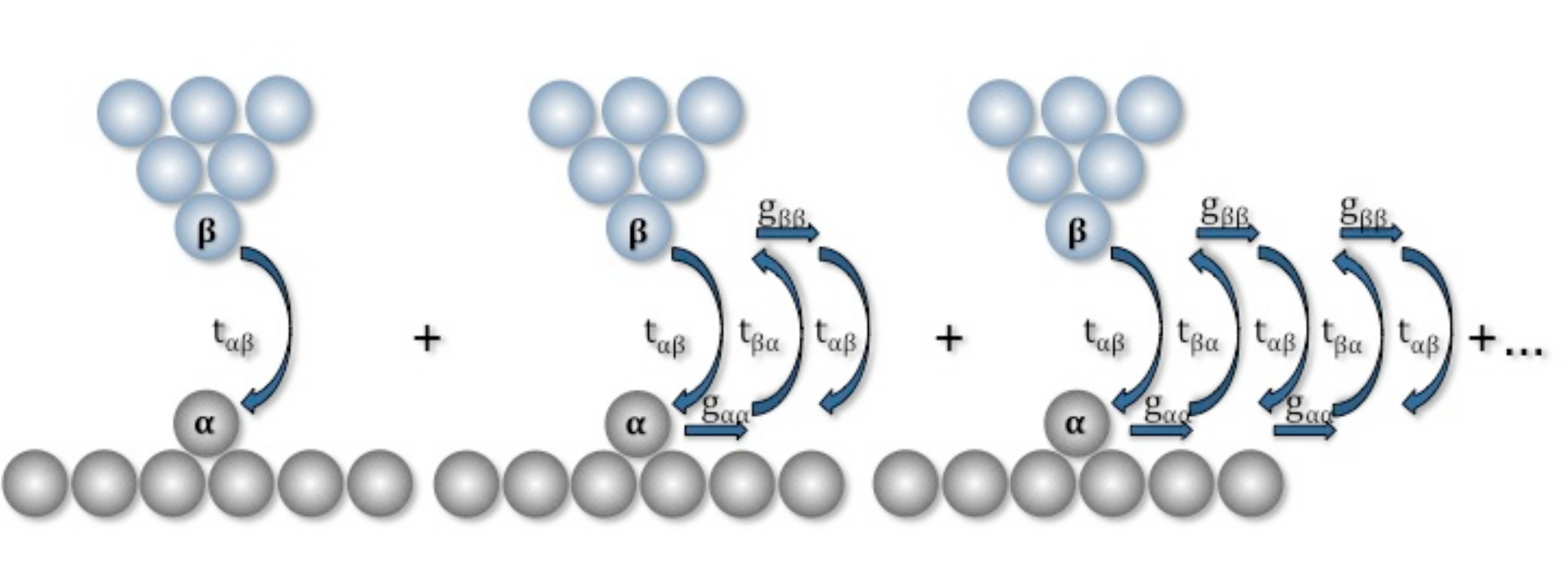}
\caption{Diagrammatic view of the electron propagation through the tunnel gap showing contributions corresponding to different orders of the perturbation theory, up to the 3rd order. }\label{Fig:scheme_tunel}
\end{figure}

When the two systems are brought into a close contact and the short-range chemical bond between the tip and the sample is formed, 
the wave functions of both systems are strongly modified. 
In this regime, a more sophisticated description of the electron transfer going beyond perturbation theory is required. 
Usually, the formulation in terms of Green's functions is the method of choice \cite{Datta05}, where all multiple scattering events of 
an electron in the contact region are involved (see fig.~\ref{Fig:scheme_tunel}).
The conductance between the tip and sample at an energy level $\epsilon$ can be described within the Landauer \cite{Landauer70} formalism as follows: 
\begin{equation}
G(\epsilon_F) = \frac{2e^2}{h} \mathrm{Tr} \left( \tau(\epsilon_F)\tau^+(\epsilon_F) \right) =   \frac{2e^2}{h}T(\epsilon_F),
\label{eq:G1}
\end{equation}
where $T$ is the transmission function giving an electron propagation probability between the tip and sample at an energy $\epsilon$ and $\tau$ and $\tau^+$ define the transfer matrix 
from the left to the right electrode and vice versa. Usually the transmission function is evaluated using the combination of ab initio calculations in a~local orbital
basis and the Green's function methods (see e.g.~\cite{Datta05, TransSiesta, Smeagol}). In a previous works \cite{JelinekSS04, MingoPRB96,Blanco06}, we have shown that the conductance can be expressed in terms of an~effective coupling between the tip $\alpha$ and the sample $\beta$ such as:
\begin{equation}
G(\epsilon_F) = \frac{8\pi^2e^2}{h} \mathrm{Tr} \left( t^{\mathit{eff}}_{\alpha,\beta}\rho_{\beta}t^{\mathit{eff}}_{\beta,\alpha}\rho_{\alpha}  \right).
\label{eq:Geff}
\end{equation}
We can express the effective coupling $t^{\mathit{eff}}$ in a diagrammatic form (see fig.~\ref{Fig:scheme_tunel}):  
$$
t^{\mathit{eff}}_{\alpha,\beta}(\epsilon_F) =  t_{\alpha,\beta} + t_{\alpha,\beta} g^r_{\beta\beta}(\epsilon_F) t_{\beta,\alpha} g^r_{\alpha\alpha}(\epsilon_F)t_{\alpha,\beta} + 
t_{\alpha,\beta} g^r_{\beta\beta}(\epsilon_F) t_{\beta,\alpha} g^r_{\alpha\alpha}(\epsilon_F)t_{\alpha,\beta}g^r_{\beta\beta}(\epsilon_F) t_{\beta,\alpha} g^r_{\alpha\alpha}(\epsilon_F)t_{\alpha,\beta} + ... 
$$
\begin{equation}
= \left[ 1- t_{\alpha,\beta} g^r_{\beta\beta}(\epsilon_F) t_{\beta,\alpha} g^r_{\alpha\alpha}(\epsilon_F) \right]^{-1}t_{\alpha,\beta}
\label{eq:Tab}
\end{equation}
and similarly
\begin{equation}
t^{\mathit{eff}}_{\beta,\alpha}(\epsilon_F)  = \left[ 1- t_{\beta,\alpha} g^a_{\alpha\alpha}(\epsilon_F) t_{\alpha,\beta} g^a_{\beta\beta}(\epsilon_F) \right]^{-1}t_{\beta,\alpha},
\label{eq:Tba}
\end{equation}
where $t_{\alpha,\beta}, t_{\alpha,\beta}$ are probabilities that an~electron will jump between tip ($\alpha$) and sample ($\beta$) and vice versa (see fig.~\ref{Fig:scheme_tunel}) 
and $g^{r,a}$ is the retarded/advanced Green's function of the uncoupled tip or sample. 

Eq.~(\ref{eq:Geff}) can be further simplified. First, we will consider the electron transport at the Fermi level $\epsilon_F$ as this is close to the conditions of 
experimental measurements on metal surfaces \cite{Ternes11}. The expressions in eqs.~(\ref{eq:Tab}) and (\ref{eq:Tba}) can be further simplified adopting a so-called wide band approximation. 
In this approximation, the real part of the Green's function tends to zero and only the imaginary part of the Green's function dominates. 
Therefore, using a~relation $\rho_{\alpha} = \frac{1}{\pi} \Im g_{\alpha\alpha}$, we can recast the expression of the conductance as follows: 
\begin{equation}
G(\epsilon_F) = \frac {8e^2}{h}\frac{\pi^2 t^2_{\alpha,\beta} \rho_{\alpha}(\epsilon_F) \rho_{\beta}(\epsilon_F)  }{\left( 1 + \pi^2  t^2_{\alpha,\beta} \rho_{\alpha}(\epsilon_F) \rho_{\beta}(\epsilon_F)  \right)^2}.
\label{eq:T2}
\end{equation}
Now, the transmission function is a function of three variables: the hopping $t_{\alpha,\beta}$ and the density of states of the tip $\rho_\beta$ 
and the sample $\rho_{\alpha}$. Next, we will find suitable expressions for these three parameters so as to get more insight into the dependence 
of the transmission function $T$ on them. 
Based on our previous arguments about the exponential decay of the wave function and its strong spatial localization, 
we will restrict ourselves to the electron transport between the surface adatom and the tip apex. 

%% hopping
In order to provide the detailed analysis of the transmission function dependence on different factors, we need to define the hopping elements between the frontier orbitals of the tip and
sample.  Several authors \cite{Harrison99, Chen93} made an attempt to find an analytical formula for the hopping elements. 
Here, we use the formula with the exponential dependence on distance proposed by J.~M.~Blanco et al.~\cite{Blanco06}:
\begin{equation}
t_{\alpha,\beta}(z) = \frac{1}{z^{m}}e^{-z\sqrt{2w_o}}, 
\label{eq:hopping}
\end{equation}
where $z$ is the distance between the tip and sample, the exponent $m$ depends on orbital symmetry associated with 
the angular momentum quantum number (for a~detailed discussion see \cite{Blanco06}) and the variable $w_o$ means the apparent height of the barrier.  
 
\begin{figure}
\includegraphics[width=\columnwidth]{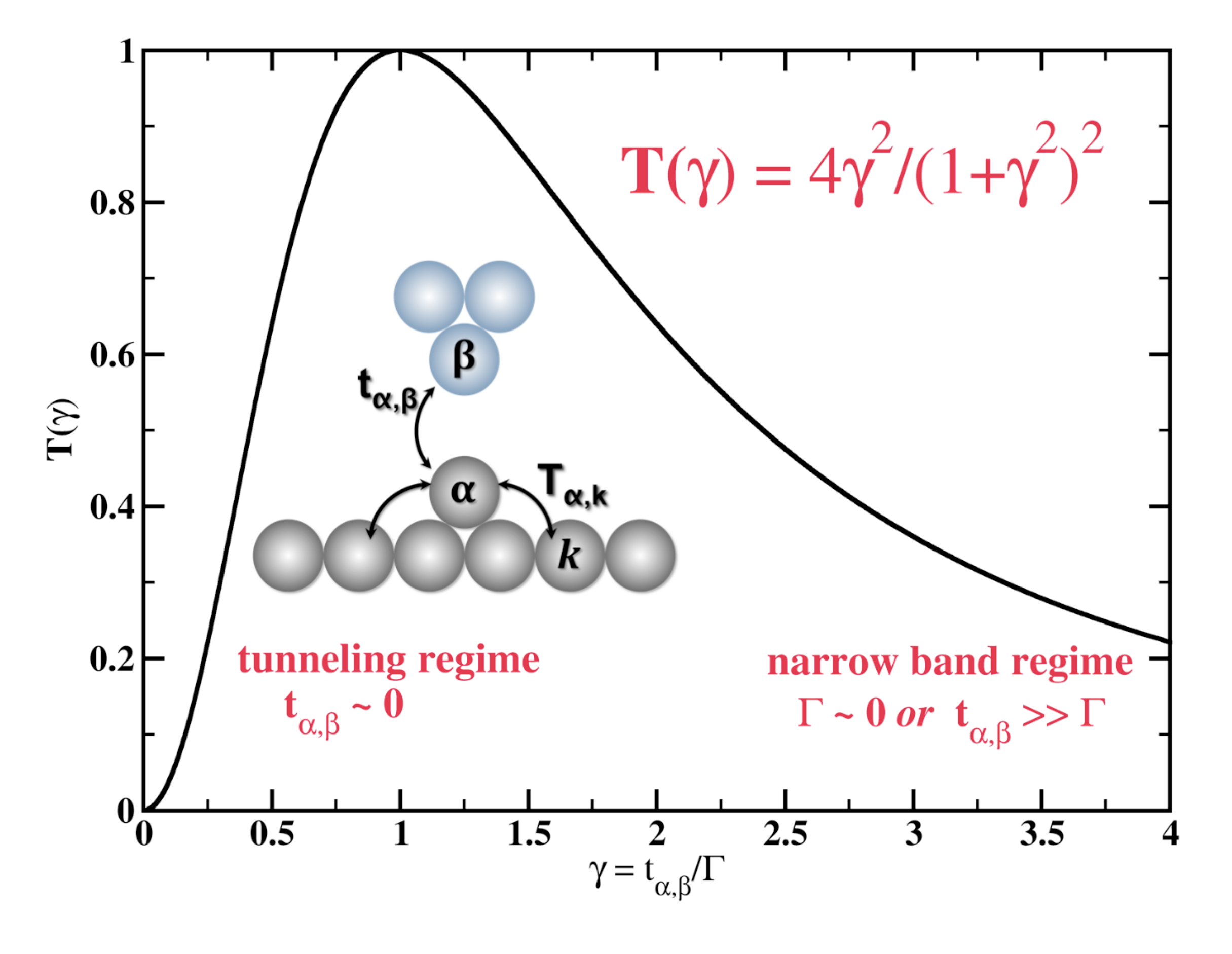}
\caption{Dependence of the transmission function $T$ on the parameter $\gamma$, which is the ratio between the coupling parameter (or reciprocal density of states) $\Gamma$ and 
the hopping $t_{\alpha,\beta}$. The inset shows a schematic view of a surface adatom coupling to the surface. The interaction represented by the hopping elements $T_{\alpha,k}$ occurs via the localized state $\alpha$ on the adatom and a surface band~$k$.}
\label{Fig:T-g}
\end{figure}

%% DOS surface adatom
We find an expression for the surface adatom density of states $\rho_{\alpha}$ via coupling of the adatom to the substrate 
(as shown in the inset of fig.~\ref{Fig:T-g}):
\begin{equation}
\rho(\epsilon) = \frac{1}{\pi} \left( \frac{\Gamma} {(\epsilon-\epsilon_{\alpha})^2+ \Gamma^2}\right), 
\label{eq:rho1}
\end{equation}
where $\Gamma$ describes a~coupling of the electronic state $\alpha$ located on the surface adatom to electronic states of the substrate 
represented by a~band {\it k}  (see inset fig. \ref{Fig:T-g}). The coupling term $\Gamma$ can be written as:
\begin{equation}
\Gamma = \sum_{k}T^2_{\alpha,k}g_k(\epsilon_F) \approx \pi \sum_{k}T^2_{\alpha,k}\rho_k(\epsilon_F).
\label{eq:Gamma1}
\end{equation}
In other words, the variable $\Gamma$ defines the energy width of the $\alpha$ state.
Finally, using eqs.~(\ref{eq:Gamma1}), (\ref{eq:rho1}) and the relation $\epsilon \approx \epsilon_{\alpha}$, we come to a~simple expression of the adatom density of states $\rho_\alpha$ in this form:  
\begin{equation}
\rho_{\alpha}(\epsilon) \approx \frac{1}{\pi\Gamma}.
\label{eq:rho2gamma}
\end{equation}

%% DOS tip apex
To express density of state of tip apex, for the sake of simplicity, we will consider a~limit case where the density of states of the tip apex and  surface adatom are similar, i.e.\ $\rho_{\beta} \approx \rho_{\alpha}$. This assumption allows us to express the transmission as a~function of a~single parameter $\gamma$. Where the parameter $\gamma$ is a~ratio between the hopping $t_{\alpha,\beta}$ and the width of the electronic state $\Gamma$
\begin{equation}
\gamma =  \frac{t_{\alpha,\beta}}{\Gamma}.
\label{eq:kappa}
\end{equation}
Using eqn. \ref{eq:kappa} the transmission coefficient $T$ can be expressed in terms of $\gamma$ as:
\begin{equation}
T(\epsilon) =  \frac{4\gamma^2}{\left(1+\gamma^2\right)^2}.
\label{eq:T3}
\end{equation}
Furthermore, to keep our model as instructive as possible we adopt another condition. We will consider only one valence 
state for each system, represented by a single {\it s}-like atomic orbital state. We also omit any modification of the surface density of states, i.e., we consider $\rho_{\alpha}$ 
or  $\rho_{\beta}$  to be independent of the energy.
Hereafter, $\gamma$ is only a number depending on the tip-sample distance via the hopping formula eq.~(\ref{eq:hopping}) and the behavior 
of the transmission function $T$ can be easily studied. Figure \ref{Fig:T-g} shows the dependence of the transmission function $T$ on the ratio $\gamma$ 
between the hopping $t_{\alpha,\beta}$ and the surface band width $\Gamma$. The optimal condition for a transmission of an electron is achieved when 
the ratio $\gamma$ tends to one, i.e.\ $t_{\alpha,\beta} \approx \Gamma$. In addition, we can identify two limiting cases when the transmission 
function diminishes: (i) the tunneling regime in which $\gamma  \rightarrow 0$ as $t_{\alpha,\beta} \rightarrow 0$ and (ii) the narrow band limit when
the coupling parameter $\Gamma$ is negligible compared to the hopping $t_{\alpha,\beta}$.
 
\begin{figure}
\includegraphics[width=\columnwidth]{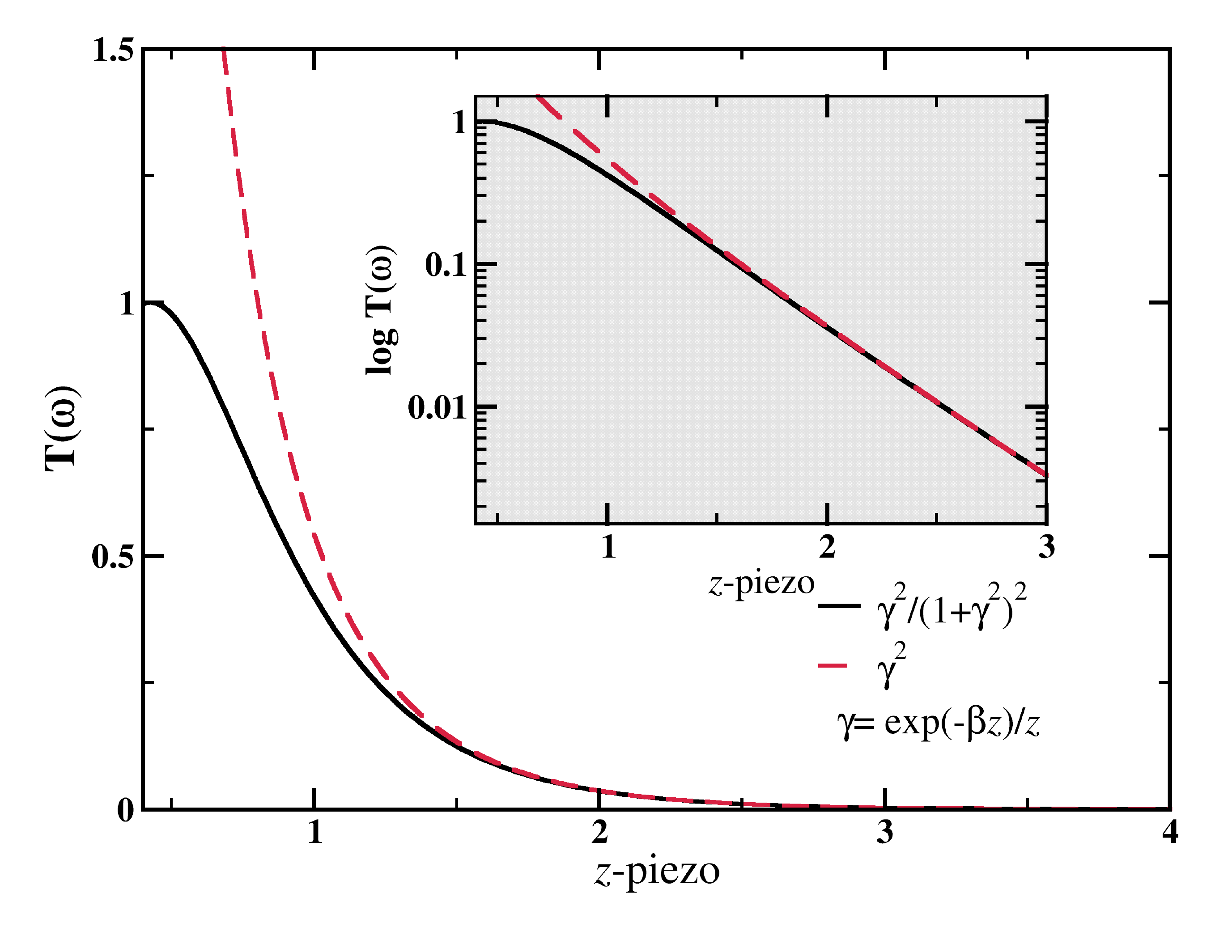}
\caption{The figure shows the dependence of the tunneling current on the tip-sample distance according to the Bardeen and the multiple-scattering theory. The inset shows the same $T$-$z$ plot but in logarithmic scale, revealing the exponential dependence of the transmission function, or current, on the tip-sample distance.}
\label{Fig:T-hop}
\end{figure}

On the other hand, the effect of the multiple scattering  on the transmission function $T$ can be demonstrated by comparing the traditional Bardeen approach,
corresponding to the 2$^{\mathrm{nd}}$ order perturbation theory, with an~exact solution that includes an~infinite series of scattering events. 
The numerator in eq.~(\ref{eq:T2}) corresponds to the Bardeen 2$^{\mathrm{nd}}$ order term, while the denominator stems from a~sum of the infinite series of 
scattering events; see eqs.~(\ref{eq:Tab}) and (\ref{eq:Tba}). Immediately, we see that the relation $T=4\gamma^2$ corresponds to eq.~(\ref{eq:T3}) up to the 2$^{\mathrm{nd}}$ order.
In other words, fig.~\ref{Fig:T-hop} reveals the difference between the Bardeen theory and the exact solution. In far distances, both curves  coincide, having 
an~exponential dependence on distance. However, at shorter distances, the Bardeen theory follows the exponential form  tending unphysically to infinity. 
In contrast, the exact solution converges to the ballistic regime with unity transmission $T=1$.
   
The discussion presented above is only a very basic picture of the electron transport between the tip and sample. One should always keep in mind, 
that there are other parameters affecting the electron transfer through the tunneling barrier. These include:
\begin{enumerate}
\item contraction of the piezo distance between the tip and sample due to atomic relaxation induced by a mutual tip-sample interaction  \cite{Blanco_PRB04,Sugimoto_PRB06},
\item change of the local density of states on both the tip apex and the surface atoms due to formation of the chemical bond between them \cite{Jelinek_PRL08,Neel07},
\item modification of the tunneling barrier (see e.g.~\cite{Lang88,Chen91a, Ternes11}),
\item modification of the surface dipole \cite{ChoPRL07,BevanPRB08}, to which the origin of the atomic contrast in Kelvin Force Probe Microscopy on semiconductor surfaces has been recently attributed~\cite{Sadewasser09}. 
\end{enumerate}

\section{The interaction energy}

In this section we describe the interaction energy arising from an interaction
of two weakly interacting systems. In general, the total energy Hamiltonian can be expressed as follows:
\begin{equation}
H = H_{\alpha}+H_{\beta}+\delta V.
\label{eq:hamtot}
\end{equation}
The Hamiltonian in eq.~(\ref{eq:hamtot}) can be written using a second
quantization formalism:
\begin{equation}
H = \sum_{\sigma}(\epsilon_{\alpha}n_{\alpha,\sigma}+\epsilon_{\beta}n_{\beta,\sigma})+t_{\alpha,\beta}\sum(c_{\alpha,\sigma}^{+}c_{\beta,\sigma}+c_{\beta,\sigma}^{+}c_{\alpha,\sigma}),
\end{equation}
where $\epsilon_{\alpha}$ (a~similar equation holds for $\epsilon_{\beta}$):
\begin{equation}
\epsilon_{\alpha}=\int\psi_{\alpha}\left(-\frac{\hbar^{2}}{2m_{e}}\nabla^{2}+V_{\alpha}(\vec{r}-\vec{R}_{\alpha})+V_{\beta}(\vec{r}-\vec{R}_{\beta})\right)\psi_{\alpha}d\vec{r}
\label{eq:e_aa}
\end{equation}
and the (real-valued) coefficients $t_{\alpha\beta}=t_{\beta\alpha}$:
\begin{equation}
t_{\alpha\beta}=\int\psi_{\alpha}\left(-\frac{\hbar^{2}}{2m_{e}}\nabla^{2}+V_{\alpha}(\vec{r}-\vec{R}_{\alpha})+V_{\beta}(\vec{r}-\vec{R}_{\beta})\right)\psi_{\beta}d\vec{r}.
\label{eq:t_ab}
\end{equation}

Using eq.~(\ref{eq:t_ab}) and (\ref{eq:e_aa}) introduced above, the interaction Hamiltonian in the orthogonal basis set $\psi$ can be written in the following matrix form: 
\begin{equation}
H=
\left( \begin{array}{cc}
h_{\alpha\alpha} & h_{\alpha\beta}\\
h_{\beta\alpha} & h_{\beta\beta}
\end{array}\right) 
=
\left( \begin{array}{cc}
\epsilon_{\alpha} & t_{\alpha\beta}\\
t_{\beta\alpha} & \epsilon_{\beta}^{o}
\end{array}\right).
\end{equation}

Let us assume that the total energies and wave-functions of the decoupled
systems $\alpha,\beta$ are known. To keep our model instructive, we will only consider the
one-electron contribution stemming from the mutual interaction between the two systems.
Hereafter, it is convenient to express the total Hamiltonian
in terms of the original \emph{atomic} wave-functions $\phi_{\alpha},\phi_{\beta}$ of the decoupled systems (here superscript
0 denotes that the matrix elements are evaluated using \emph{atomic}
orbitals on each subsystem $\alpha, \beta$, which are not yet orthogonal):
\begin{equation}
H^{o}=
\left( \begin{array}{cc}
h^{o}_{\alpha\alpha} & h^{o}_{\alpha\beta}\\
h^{o}_{\beta\alpha} & h^{o}_{\beta\beta}
\end{array}\right) 
=
\left( \begin{array}{cc}
\epsilon_{\alpha}^{o}+\delta h^{o}_{\alpha\alpha} & t^{o}_{\alpha\beta}\\
t^{o}_{\beta\alpha} & \epsilon_{\beta}^{o}+\delta h^{o}_{\beta\beta}
\end{array}\right),
\label{eq:Ho}
\end{equation}
where the terms $\delta h^{o}_{\alpha\alpha}$ and $\delta h^{o}_{\beta\beta}$ mean a small change of the \emph{atomic} levels $\epsilon_{\alpha}^{o}$, $\epsilon_{\beta}^{o}$ of the subsystems $\alpha,\beta$, defined by eqs.~(\ref{eq:se1}), (\ref{eq:se2}), respectively.
Matrices $h^{o}_{\alpha\beta}$, $h^{o}_{\beta\alpha}$ denote hopping elements between the two subsystems $\alpha,\beta$. Note that terms $h^{o}_{\alpha\alpha}$ and $t^{o}_{\beta\alpha} $ in eq.~(\ref{eq:Ho}) are defined as :
%% diagonal correction
\begin{equation}
h^{o}_{\alpha\alpha}=\int\phi_{\alpha}\left(-\frac{\hbar^{2}}{2m_{e}}\nabla^{2}+V_{\alpha}(\vec{r}-\vec{R}_{\alpha})+V_{\beta}(\vec{r}-\vec{R}_{\beta})\right)\phi_{\alpha}d\vec{r},
\label{eq:hoaa}
\end{equation}
%% hopping
and the hopping term $t_{\alpha\beta}^{o}$:
\begin{equation}
t^{o}_{\alpha\beta}=\int\phi_{\alpha}\left(-\frac{\hbar^{2}}{2m_{e}}\nabla^{2}+V_{\alpha}(\vec{r}-\vec{R}_{\alpha})+V_{\beta}(\vec{r}-\vec{R}_{\beta})\right)\phi_{\beta}d\vec{r}.
\end{equation}
Comparing eqs.~(\ref{eq:se1}) and (\ref{eq:hoaa}) with eq.~(\ref{eq:Ho}), we obtain  directly an expression for $\delta h^{o}_{\alpha\alpha}$:
\begin{equation}
\delta h^{o}_{\alpha\alpha}=\int\phi_{\alpha} V_{\beta}(\vec{r}-\vec{R}_{\beta}) \phi_{\alpha}d\vec{r}.
\label{eq:dhoaa}
\end{equation}
It is evident that the integral vanishes if the potential $V_{\beta}(\vec{r}-\vec{R}_{\beta})$ is short-ranged. Therefore, without presence of a long-range potential (e.g.\ a dipolar coulombic interaction) we can put:
 \begin{equation}
 \delta h^{o} = 0 \Rightarrow  \ \epsilon \approx \epsilon^{o}.
 \label{eq:eeqe0}
 \end{equation} 
It is worth commenting that the term $\delta h^{o}$ represents the first order term in the perturbation theory.
Non-orthogonality of the \emph{atomic} wave functions  $\phi_{\alpha}, \phi_{\beta}$  gives rise to a global matrix overlap $\Theta=I+S$, where 
\begin{equation}
S=\left(\begin{array}{cc}
0 & S_{\alpha\beta}\\
S_{\beta\alpha} & 0\end{array}\right).
\label{eq:S}
\end{equation}

The new (orthonormal) basis $\psi_{i}\,$ is orthogonalized with both systems
through the L\"{o}wdin transformation \cite{Lowdin63}:
\begin{equation}
\psi_{i}=\sum_{i,j}\Theta_{ij}^{-\frac{1}{2}}\phi_{j}.
\end{equation}
The operator $\Theta\ $can be expanded up to the second order in $S\,$ as
\begin{equation}
\Theta^{-\frac{1}{2}} \approx I-\frac{1}{2}S+\frac{3}{8}S^{2}.
\label{eq:S2}
\end{equation}

Our goal is to express the orthogonal Hamiltonian $H$ of the whole system in terms of the unperturbed  {\it atomic} hamiltonian $H^{o}$. 
The transformed Hamiltonian matrix $H$ in the new orthogonal space by the $S^2$ expansion via eq.~(\ref{eq:S2}) can now be expressed as:

$$
H = \Theta^{-\frac{1}{2}} H^{o} \Theta^{-\frac{1}{2}} \approx \left(I-\frac{1}{2}S+\frac{3}{8}S^{2} \right) H^{o} \left(I-\frac{1}{2}S+\frac{3}{8}S^{2} \right) 
$$
\begin{equation}
= H^{o}-\frac{1}{2}(SH^{o}+H^{o}S)+\frac{1}{4}SH^{o}S+\frac{3}{8}(H^{o}S^{2}+S^{2}H^{o}) + O(S^3),
\label{eq:HoS2}
\end{equation}
where $H^o$ is defined by eq.~(\ref{eq:Ho}). Introducing eqs.~(\ref{eq:Ho}) and (\ref{eq:S}) into eq.~(\ref{eq:HoS2}) we can obtain, with a little algebra, the desired expression for the matrix elements of $H$. In the next sections, we will discuss the intraatomic (diagonal) and interatomic (off-diagonal) terms of the Hamiltonian.
Finally, we will find the expression for the interaction energy arising from the interaction of the two weakly interacting systems. 
To do this, we will assume the following conditions:
\begin{enumerate}
\item the atomic potentials $V_{\alpha}$ and $V_{\beta}$ do not overlap i.e. $V_{\alpha}\cdot V_{\beta}=0$,
\item three-center terms are negligible,
\item the diagonal matrix elements $h^{o}_{\alpha\alpha}$ and $h^{o}_{\beta\beta}$ in the non-orthogonal $\phi$ basis set approximately equal the atomic orbital energies $\epsilon^{o}_{\alpha\alpha}, \epsilon^{o}_{\beta\beta}$. 
\end{enumerate}
The last condition is valid if the integral $\delta h_{\alpha\alpha}^{o} = \int\phi_{\alpha}\left(V_{\beta}(\vec{r}-\vec{R}_{\beta})\right)\phi_{\alpha}d\vec{r}$
vanishes. In other words, if the potential $V_{\beta}(\vec{r}-\vec{R}_{\beta})$ is short-ranged. 

\subsection{Off-diagonal (interatomic) terms in $\mathbf{S^{2}}$ approximation}
First, we will discus the off-diagonal terms $t$. From eq.~(\ref{eq:HoS2}), we get
\begin{equation}
t_{\alpha\beta}=h_{\alpha\beta}^{o}-\frac{1}{2}S_{\alpha\beta}(h^{o}_{\alpha\alpha}+h^{o}_{\beta\beta}).
\end{equation}
If there is a negligible change of \emph{atomic} levels due to mutual interaction of systems $\alpha, \beta$, e.g.\ represented by long-range coulombic interaction, then using eq.~(\ref{eq:eeqe0}) we can write $t_{\alpha\beta}$ as:
\begin{equation}
t_{\alpha\beta}=h_{\alpha\beta}^{o}-\frac{1}{2}S_{\alpha\beta}(\epsilon^{o}_{\alpha\alpha}+\epsilon^{o}_{\beta\beta}).
\label{eq:T-S}
\end{equation}

\subsection{Diagonal (intraatomic) terms in $\mathbf{S^{2}}$ approximation}

The wave-function overlap between the subsystems gives rise to a~repulsive term $\delta^{S}h$ renormalizing the original diagonal elements of each subsystem:
\begin{equation}
\delta^{S}h_{\alpha\alpha}=-\frac{1}{2}\sum_{\beta}\left[S_{\alpha\beta}t_{\beta\alpha}+t_{\alpha\beta}S_{\beta\alpha}\right]
+\frac{1}{4}\sum_{\beta}S_{\alpha\beta}S_{\beta\alpha}\left[h^{o}_{\alpha\alpha}-h^{o}_{\beta\beta}\right],
\end{equation}
and
\begin{equation}
\delta^{S}h_{\beta\beta}=-\frac{1}{2}\sum_{\alpha}\left[S_{\beta\alpha}t_{\alpha\beta}+t_{\beta\alpha}S_{\alpha\beta}\right]
+\frac{1}{4}\sum_{\alpha}S_{\beta\alpha}S_{\alpha\beta}\left[h^{o}_{\beta\beta}-h^{o}_{\alpha\alpha}\right],
\end{equation}
and using eq.~(\ref{eq:eeqe0}) we finally obtain:
\begin{equation}
\delta^{S}h_{\alpha\alpha}= -\sum_{\beta} S_{\alpha\beta}t_{\beta\alpha} 
+ \frac{1}{4}\sum_{\beta}S^2_{\alpha\beta}\left[\epsilon^{o}_{\alpha}-\epsilon^{o}_{\beta}\right],
\end{equation}
and
\begin{equation}
\delta^{S}h_{\beta\beta}= -\sum_{\alpha} S_{\alpha\beta}t_{\alpha\beta}
+ \frac{1}{4}\sum_{\alpha}S^2_{\alpha\beta}\left[\epsilon^{o}_{\beta}-\epsilon^{o}_{\alpha}\right].
\label{eq:dhs}
\end{equation}

\begin{figure}
\includegraphics[width=\columnwidth]{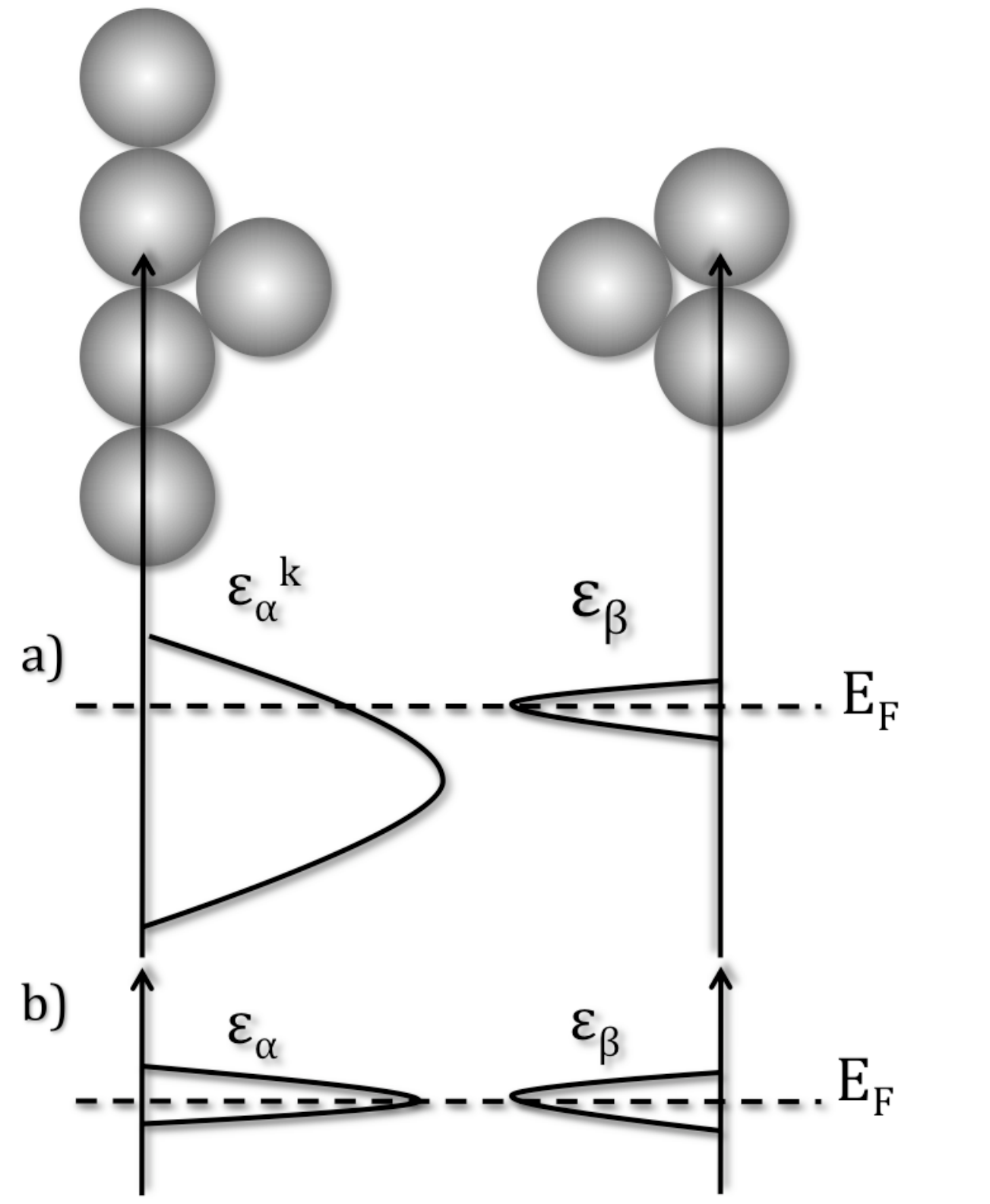}
\caption{Schematic view of two limiting cases considered in the text.  Two interacting atomic levels: (a) a strongly localized state $\epsilon_\beta$ (tip) interacting with a delocalized surface  band $\epsilon^k_\alpha$; (b) two degenerate localized states $\epsilon_\alpha$ and $\epsilon_\beta$ placed at the Fermi level.}
\label{Fig:levels_PT}
\end{figure}

\subsection{The chemical interaction}

So far, the energy interaction terms derived in the previous sections, see eqs.~(\ref{eq:T-S}) and (\ref{eq:dhs}), have introduced only the energy corrections due to orthogonalization of wave 
functions of each subsystem. For our discussion about the relation between the tunneling current  and the chemical force, we also need to take into account 
the energy contribution emerging from the formation of the chemical bond between the frontier states of the tip and sample. We will restrict ourselves to the one-electron approximation.
Therefore the effective hopping matrix element can be expressed via the Bardeen tunneling contribution (for a~detailed discussion see \cite{Goldberg89}):
\begin{equation}
t_{\alpha\beta}=T_{\alpha\beta}^{B}=-\frac{\hbar^{2}}{2m}\int_{\Omega}(\phi_{\beta}^{*}\nabla\phi_{\alpha}-\phi_{\alpha}\nabla\phi_{\beta}^{*})d\vec{s}.
\label{eq:BTC}
\end{equation}

%Note, that for the case of a~long-range potential, one can expect $t_{\alpha\beta}$ to be well approximated by the two-center  Bardeen's
%tunneling current $T_{\alpha\beta}^{B}$ \ref{eq:BTC} corrected by a~constant factor $\gamma^{LR}$ (for detailed discussion see \cite{Goldberg89}):
%\begin{equation}
%h_{\alpha\beta}=\gamma^{LR} T_{\alpha\beta}^{B}\label{eq:hop2bardeen}
%\end{equation}

Figure \ref{Fig:levels_PT} shows two limiting cases we will consider for the interaction between the single-electron states of the two subsystems:
a strongly localized state $\epsilon_\beta$ (tip) interacting with a delocalized surface band $\epsilon^k_\alpha$; (b) two degenerate localized states $\epsilon_\alpha$ and $\epsilon_\beta$ placed at the Fermi level.
In the former, the  interaction energy due to hybridization of two states is given
using the standard perturbation theory up to the second order by:
\begin{equation}
\delta E_{\alpha}^{chem}=\sum_{\beta}\frac{|h_{\alpha\beta}|^{2}}{\epsilon_{\alpha}^{o}-\epsilon_{\beta}^{o}}\approx\sum_{\beta}\frac{| T_{\alpha\beta}|^{2}}{\epsilon_{\alpha}^{o}-\epsilon_{\beta}^{o}},
\label{eq:ch2nd}
\end{equation}
\begin{equation}
\delta E_{\beta}^{chem}=\sum_{\alpha}\frac{|h_{\alpha\beta}|^{2}}{\epsilon_{\beta}^{o}-\epsilon_{\alpha}^{o}}\approx\sum_{\alpha}\frac{| T_{\alpha\beta}|^{2}}{\epsilon_{\beta}^{o}-\epsilon_{\alpha}^{o}}.
\end{equation}
It is important to note that the first order term vanishes here.
Namely, in this approximation, we do not consider any contribution
to the diagonal elements $\epsilon^{o}_{\alpha(\beta)}$ in the perturbation
potential already expressed in the new orthonormal basis. In particular,
this is an important point in our discussion of the force vs.\ current
scaling.

On the other hand, the latter case requires application of degenerate perturbation theory, where the lowest order gives:
\begin{equation}
\delta E_{\alpha}^{chem} =\sum_{\beta} {|h_{\alpha\beta}|} \approx\sum_{\beta} T_{\alpha\beta},
\label{eq:ch1th}
\end{equation}
\begin{equation}
\delta E_{\beta}^{chem}=\sum_{\alpha}{|h_{\alpha\beta}|} \approx\sum_{\alpha} T_{\alpha\beta}.
\end{equation}
An example of this limiting case can be represented by an~interaction between two strongly localized states, such as two dangling bonds being present on a~semiconductor surface as well as on the tip.
As we will see in the following discussion, it is just the character (localization) of the two interacting electronic states which has a fundamental consequence to the relation between the tunneling current and the short-range force in atomic contacts.

\section{Relation between current \& interaction energy}

In this section, we will derive the relation between the tunneling current and the force on the onset of the chemical bond formed by the frontier electron states of the tip and sample.
According to the previous discussion, we already know the tunneling current can be expressed using Fermi's golden rule as:

\begin{equation}
I_{t}\sim(T^{B})^{2}\delta(\epsilon_{\alpha}-\epsilon_{\beta}).
\label{eq:current}
\end{equation}
From this expression it is immediately evident that the tunneling current $I_t$ in the far distance regime is proportional to $(T^{B})^{2}$.
On the other hand, using eqs.~(\ref{eq:T-S}), (\ref{eq:dhs}), the interaction energy is given by:
$$
E^{int}= -\sum_{\beta}\left[S_{\alpha\beta} T_{\beta\alpha}^{B}+ T_{\alpha\beta}^{B}S_{\beta\alpha}\right]+\frac{1}{2}\sum_{\beta}S_{\alpha\beta}S_{\beta\alpha}\left[\epsilon_{\alpha}^{o}-\epsilon_{\beta}^{o}\right]-
$$
\begin{equation}
-\sum_{\alpha}\left[S_{\beta\alpha} T_{\alpha\beta}^{B}+ T_{\beta\alpha}^{B}S_{\alpha\beta}\right]+\frac{1}{2}\sum_{\alpha}S_{\beta\alpha}S_{\alpha\beta}\left[\epsilon_{\beta}^{o}-\epsilon_{\alpha}^{o}\right] + E^{chem}.
%+2\sum_{\beta}\frac{| T_{\alpha\beta}^{B}|^{2}}{\epsilon_{\alpha}^{o}-\epsilon_{\beta}^{o}}
%+2\sum_{\alpha}\frac{| T_{\alpha\beta}^{B}|^{2}}{\epsilon_{\beta}^{o}-\epsilon_{\alpha}^{o}}
\end{equation}
Finally, the one-electron total energy $E^{int}$ can be expressed as a function of $S$ and $T_{\alpha\beta}^{B}$:
\begin{equation}
E^{int}\approx S^{2}\Delta\epsilon - ST^{B} +  E^{chem},
\label{eq:intE}
\end{equation}
where the term $E^{chem}$ is proportional to $T^2$ or $T$ for non-degenerate and degenerate electronic states, respectively; see eqs.~(\ref{eq:ch1th}) and (\ref{eq:ch2nd}). 

Three terms appear in eq.~(\ref{eq:intE}). The first two terms are a~consequence
of the orthogonalization process between $\phi_{\alpha},\phi_{\beta}$ wave
functions. The first term, which shifts the diagonal elements, decays fast with
distance by $\sim S^{2}$, therefore we can consider it insignificant
at far distance. The second term is directly proportional to $ST$.
From eq.~(\ref{eq:T-S}), we see that the overlap $S$ is proportional to the hopping element $T$; 
i.e.\ $T \sim S$. Using this relation, we immediately conclude that the
first two terms (i) mostly cancel each other and (ii) are proportional to $T^{2}$. In other words, it implies the interaction energy due to the orthogonalization process  
is proportional to the tunneling current as $I\sim E$, following the Hofer-Fisher prediction \cite{Hofer03}:
\begin{equation}
E^{S2}\approx S^{2}\Delta\epsilon-ST^{B} \sim {(T^{B})}^2.
\label{eq:intS2}
\end{equation}
 
The third term in eq.~(\ref{eq:intE}), representing the level hybridization, shows an interesting behavior.
In more general cases, we have two non-degenerate and/or delocalized states, for which 
the interaction energy is described via the 2$^{\mathrm{nd}}$ order term of eq.~(\ref{eq:ch2nd}):
\begin{equation}
E^{chem}_{2nd} \approx \frac{(T^{B})^{2}}{\Delta\epsilon}.
\label{eq:intE2nd}
\end{equation}
Here, we immediately find once again that the tunneling current is directly proportional to the the interaction energy.

%% Figure 23 level model
\begin{figure}
\includegraphics[width=\columnwidth]{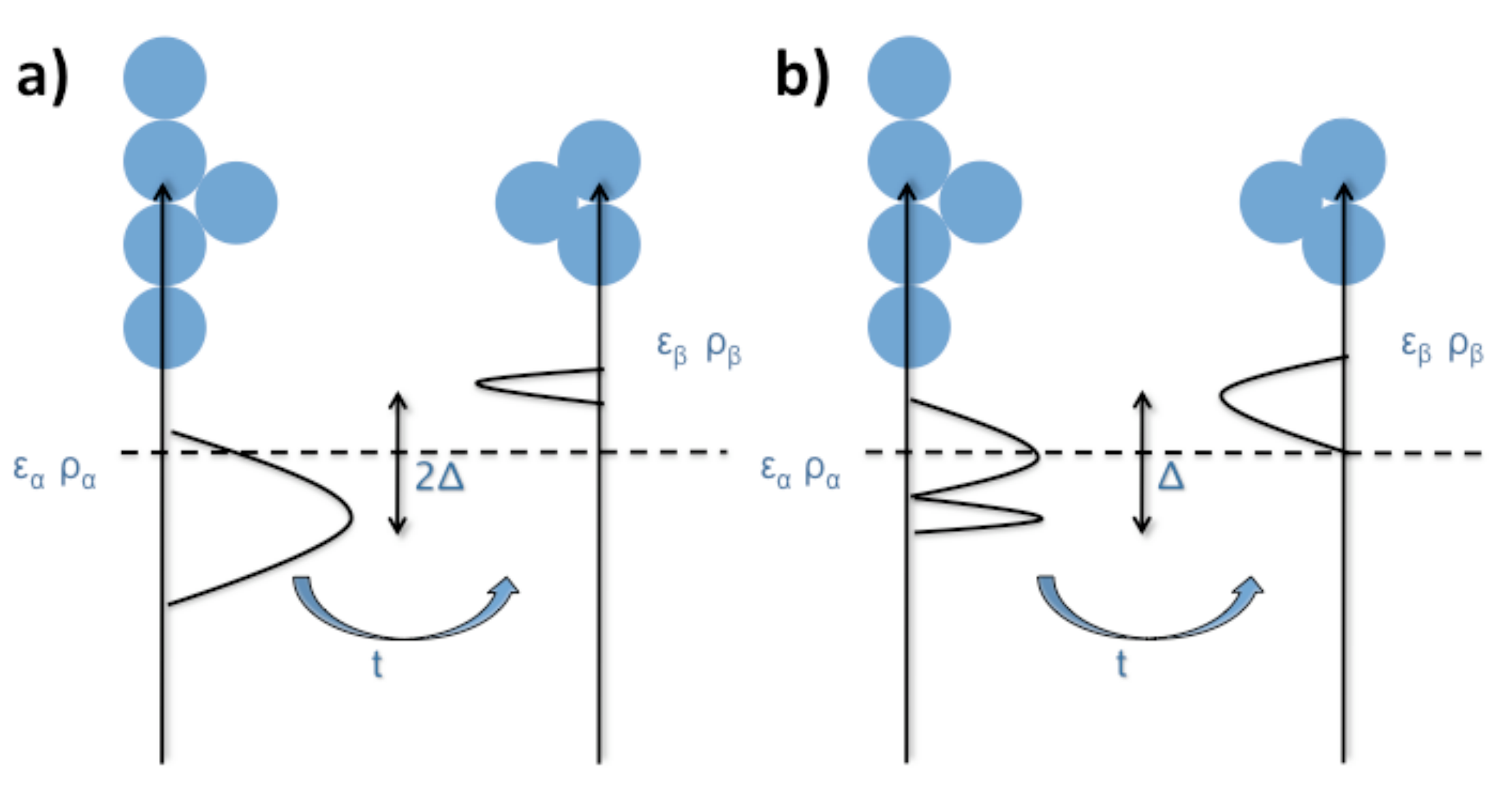}
\caption{Schematic view of two model systems with 2 (left) and 3 (right) electronic states, respectively.}\label{Fig:23level}
\label{fig:23model}
\end{figure}

Nevertheless, the situation changes if two degenerate and localized states are considered. 
Using eq.~(\ref{eq:ch1th}) we have:
\begin{equation}
E^{chem}_{1th} \approx T^{B}.
\label{eq:intE1th}
\end{equation}
Comparing eqs.~(\ref{eq:intE1th}), (\ref{eq:intS2}) and (\ref{eq:current}), we obtain a non-trivial dependence relating the current and the
force. Usually, the short-range chemical interaction $E^{chem}$ shows a slower increase at larger tip-sample distances than 
the~$E^{S2}$ energy term. Therefore with a dominance of the $E^{chem}$ term,  the relation between the current and the atomic force
follows a  $I\sim E^{2}$ power law as predicted by Chen \cite{Chen91,Chen05}. 

Based on our previous analysis, it is evident that the relation between the tunneling current and the interaction
energy depends strongly on the relative magnitudes of the three terms in our model described by eq.~(\ref{eq:intE}). 
Indeed, Hofer \& Fisher \cite{Hofer03} neglected the first order term in the interaction energy, justifying
this step by the fact that both wave functions and potentials of the tip
and sample decay exponentially and the perturbation potential of
tip's wave function is the surface potential and vice versa. This
makes the perturbation of the diagonal elements negligible.

\subsection*{Numerical models}

To gain more insight into the relation between the tunneling current and the interaction energy as a~function of the tip-sample distance, number of states and their energy alignment, 
we will consider a~simple numerical model (see Fig.~\ref{Fig:23level}). The model consists of two subsystems $\alpha,\beta$ defined by molecular levels
with energies $\epsilon_\alpha$ and $\epsilon_\beta$ and coupled by the hopping elements rewritten from eq.~(\ref{eq:hopping}) as:
\begin{equation}
t_{\alpha\beta} = \frac{1}{z^m}e^{-\lambda z},
\label{eq:hopmodel}
\end{equation} 
where $m = l_{\alpha} + l_{\beta} + 1$, $l$ is the angular momentum quantum number of a given state $\alpha$ or $\beta$, $z$ is the distance between the tip and sample and 
$\lambda$ is the characteristic decay constant related to the effective work function of the system via $\lambda \approx 2\sqrt{w_o}$.  
The electron occupancy of individual levels is defined according to their position with respect to the Fermi level (plotted as a~dashed line in Fig.~\ref{Fig:23level}). It means (i) the states bellow the Fermi level are fully occupied, (ii) the states on the Fermi level are half-filled and (iii) the states above the Fermi level are empty. Then the interaction energy is defined as a~difference between sum of occupied states multiplied by their occupancy of coupled and decoupled systems.
  
In the case of a two-level system including only $s$-like electronic states 
on both subsystems (see fig.~\ref{Fig:23level}), the energy of the coupled system is determined by the secular equation of the following matrix:
\begin{equation}
\left(\begin{array}{cc}
\epsilon_{\alpha}-\epsilon & t_{s-s}\\
t_{s-s} & \epsilon_{\beta}-\epsilon\end{array}\right).
\label{eq:matH-2L}
\end{equation} 
On the other hand, the density of states $\rho_\alpha$ of the $\alpha$ state is given by:
\begin{equation}
\rho_{\alpha}(\epsilon)=\frac{1}{\pi}\frac{\Gamma_{\alpha}}{\left[(\epsilon-\epsilon_{\alpha})^{2}+\Gamma_{\alpha}^2\right]},
\label{eq:rho}
\end{equation} 
where $\Gamma_\alpha$ is the coupling constant to an electrode (see discussion in the section \ref{sec:SIR}). 
Then the energy of a given electronic level $\alpha$ is written as (see fig. \ref{fig:23model}): 
\begin{equation}
\epsilon_{\alpha}=-\Delta/2 - i\ \Gamma_{\alpha}.
\label{eq:estatea}
\end{equation}
Similarly, the energy of the electronic state $\epsilon_\beta$ is evaluated as:
\begin{equation}
\epsilon_{\beta}=+\Delta/2 - i\ \Gamma_{\beta}.
\label{eq:estateb}
\end{equation}  

Here we restrict our discussion to distances at which only a~weak chemical force forms between the tip and the sample. Therefore, we can neglect a~change of the density of states 
given by eq.~(\ref{eq:rho}) and the contraction of the tip-sample distance. As we have seen in the previous analysis, in this interaction regime the Bardeen approximation is still valid 
(see fig.~\ref{Fig:T-hop}).  Then the conductance at the Fermi energy $\epsilon_F$ between two levels $\epsilon_{\alpha,s}$ and $\epsilon_{\beta,s}$ can be expressed using eq.~(\ref{eq:current}) as:
\begin{equation}
G_{t}(\epsilon_F) = \frac{4\pi e}{\hbar}Tr(\rho_\alpha(\epsilon_F) t_{\alpha,\beta} \rho_\beta(\epsilon_F) t_{\beta, \alpha}).
\label{eq:I2L}
\end{equation}
Solving the secular equation for the two-level system, eq.~(\ref{eq:matH-2L}), we obtain the expression for the interaction energy (where $t$ means $t_{s-s}$):
 \begin{equation}
% Preview source code for paragraph 69
E_{1,2}=\frac{-i\left(\Gamma_{\alpha}+\Gamma_{\beta}\right)}{2}\pm\frac{\sqrt{-(\Gamma_{\alpha}+\Gamma_{\beta})^{2}+4(t^{2}+\Delta+\Gamma_{\alpha}\Gamma_{\beta}-Im\Delta(\Gamma_{\beta}-\Gamma_{\alpha}))}}{2}.
\end{equation}
If the coupling of both levels is equal, i.e.\ $\Gamma_\alpha = \Gamma_\beta$, the interaction energy becomes even more simple:
\begin{equation}
E_{1,2} =\frac{i \Gamma}{2} \pm \sqrt{t^{2}+\Delta^{2}},
\label{eq:solution2L}
\end{equation}
and immediately, we see that the interaction energy is proportional to $t$  if the condition $t \gg \Delta$ is satisfied so that $\sqrt{t^2 + \Delta^2} = t$. By direct comparison with eq.~(\ref{eq:I2L})
we find the relation between the interaction energy $E$ and the current $I_t$ to be $I_t \sim E^2$. 
In the other limiting case, when the relation $t < \Delta$ is satisfied,  we use $\sqrt{t^2 + \Delta^2} = \Delta + \frac{t^2}{2\Delta} - \frac{t^4}{8\Delta^3} + ...$ to obtain $E \sim t^2$.
Therefore in this interaction regime, the current and the interaction energy (force) are proportional, $I_t \sim E$. In conclusion,  we have found two regimes according to which the interaction energy
may be proportional to the current, $I_t \sim E$, or have a~quadratic relation with it, $I_t \sim E^2$, in good agreement with our findings in the previous section.

\begin{figure}
\includegraphics[width=\columnwidth]{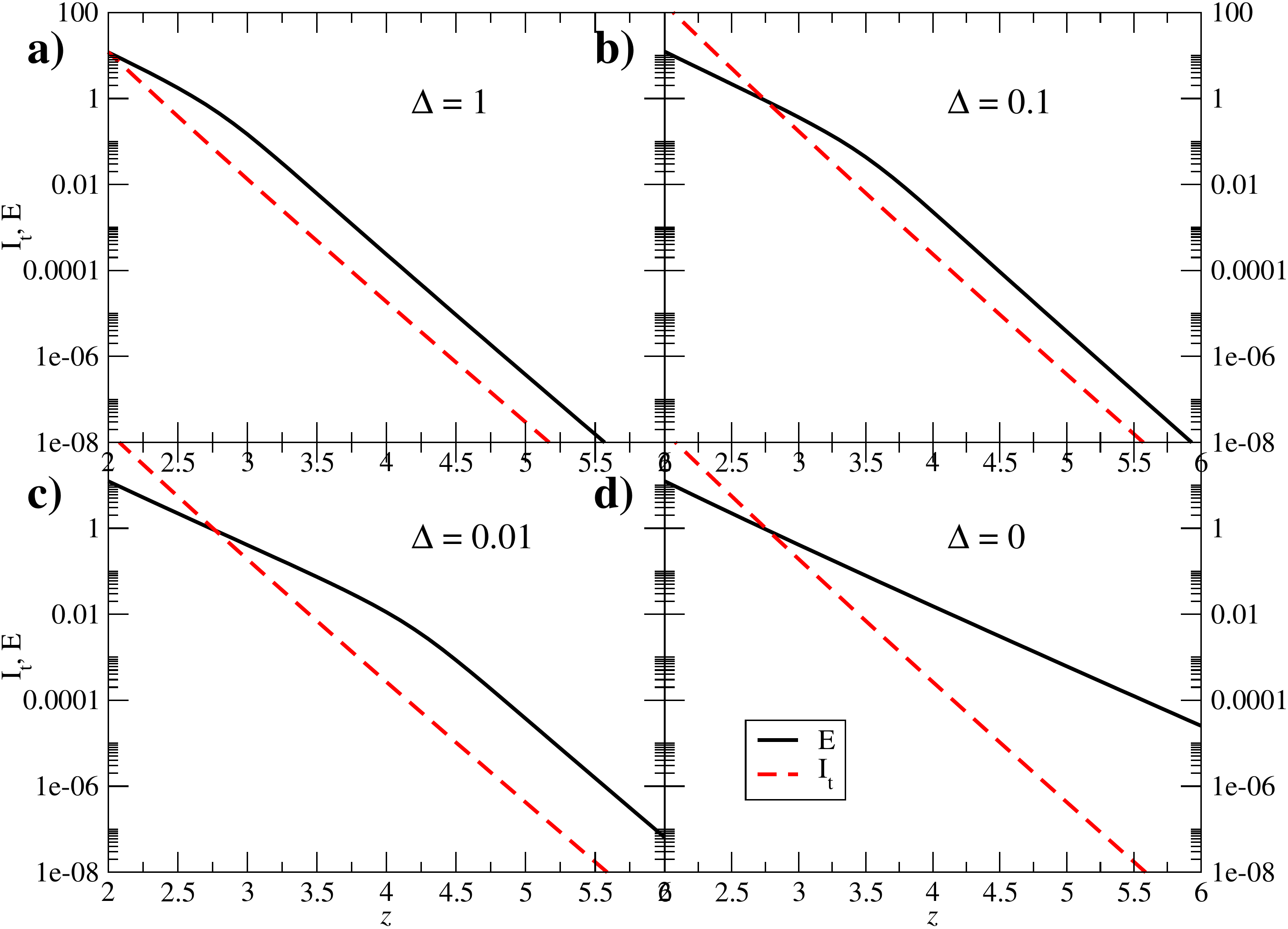}
\caption{Logarithmic plots of the tunneling current $I_t \times \left(\frac{4\pi e}{\hbar} U \right)^{-1}$ and the interaction energy $E$ along the distance for the 2-level system (see fig.~\ref{Fig:23level} on the left). 
}\label{Fig:RES2L}
\end{figure}

A~more generalized picture, including the dependence on the distance $z$ and the energy alignment $\Delta$, can be obtained by a~direct diagonalization of the matrix given 
by eq.~(\ref{eq:matH-2L}). The results for different values of the energy alignment $\Delta$ of two $s$-like states $\epsilon_{\alpha,s}$ and $\epsilon_{\beta,s}$ are shown in fig.~\ref{Fig:RES2L}, 
where the interaction energy and the tunneling current, the latter given by eq.~(\ref {eq:I2L}), are plotted as a~function of the distance $z$. In this particular case, the hopping element is written 
according to eq.~(\ref{eq:hopping}) as $t_{\alpha\beta} = \frac{A_o}{z}e^{-z\lambda}$, where we used the parameters  $\lambda = 2.828$ and $A_o = 10^3$.
The density of states is obtained from eq.~(\ref{eq:rho}) with the coupling constant $\Gamma=0.3$ for both states $\alpha$ and $\beta$.
The logarithmic plots of the interaction energy $E$ and the tunneling current  $I_t$ at the Fermi level with different degeneracy $\Delta$ of states $\epsilon_\alpha$ 
and $\epsilon_\beta$ reveals the exponential dependence of both variables on the distance.  For the sake of simplicity we considered the Fermi level $\epsilon_F = 0$ and we omitted 
the constant $\left(\frac{4\pi e}{\hbar}\right)$ and the voltage ($U$) dependence when plotting the current $I_t$. The results are shown in fig.~\ref{Fig:RES2L}. 
Comparing figs.~\ref{Fig:RES2L} a) to d) we can clearly distinguish the two scaling regimes $I_t \sim E$ or $I_t \sim E^2$, respectively. 
The existence of a given regime is controlled by the ratio between the degeneracy parameter $\Delta$ and the hopping $t_{\alpha\beta}$. 
Note, in the case of the fully degenerate electronic levels (see fig.~\ref{Fig:RES2L}d)) only the quadratic relation $I_t \sim E^2$ occurs.

%% Figure 23 level model

%3-level model
In the next step, let's consider a~more complex case including 3~levels $\epsilon_{\alpha,s}$, $\epsilon_{\alpha,d}$ and $\epsilon_{\beta,s}$ (see fig.~\ref{Fig:23level}b) keeping 
parameters $\lambda$, $\Gamma$ and $A_o$ the same as in the 2-level model. The model mimics surfaces where both $s$-like and $d$-like states are present, such as transition 
and noble metal surfaces. The interaction Hamiltonian has the form:
\begin{equation}
\label{eq:matH-3L}
\left(\begin{array}{ccc}
\epsilon_{\alpha,s}-\epsilon & 0 & t_{s-s}\\
0 & \epsilon_{\alpha,d}-\epsilon & t_{s-d}\\
t_{s-s} & t_{s-d} & \epsilon_{\beta,s}-\epsilon\end{array}\right),
\end{equation} 
where the energy levels are given by eq.~(\ref{eq:estatea}), (\ref{eq:estateb}). The tunneling current is also evaluated using eq.~(\ref{eq:I2L}) as in the case of the 2-level model.

\begin{figure}
\includegraphics[width=\columnwidth]{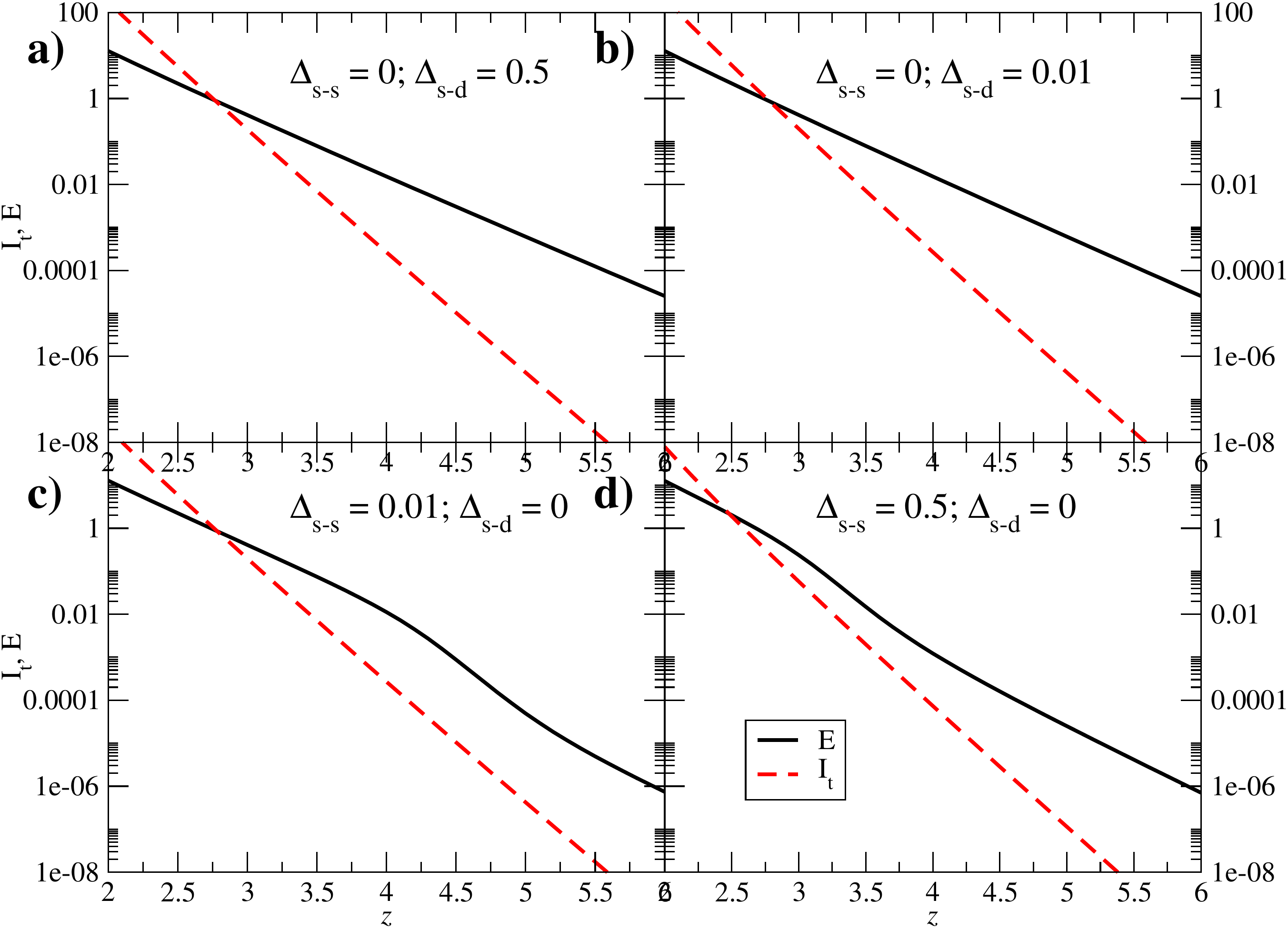}
\caption{Logarithmic plots of the tunneling current $I_t \times \left(\frac{4\pi e}{\hbar} U \right)^{-1}$ and the interaction energy $E$ along the distance for the 3-level system 
(see fig.~\ref{Fig:23level} on the right).  An~influence of the degeneracy state parameter $\Delta$  on the relation between $I_t$ and $E$ is seen. 
The two characteristic scaling regimes $I_t \approx E$  or $I_t \approx E^2$ are found depending on the degeneracy $\Delta$ and the hopping $t_{s,x}\ (x=s,d)$ controlled by the distance 
$z$.
}\label{Fig:RES3L}
\end{figure}

The dependence of both the current $I_t$ and the energy $E$ on the distance is shown in fig.~\ref{Fig:RES3L}. This model gives the characteristic scaling regimes close to $I_t \sim E$ 
or $I_t \sim E^2$ again, but with more intricate occurrence along the distance $z$. From eq.~(\ref{eq:hopmodel}), it is clear that the hopping element $t_{s-s}$ between 
two $s$-like orbitals has slower decay than the hopping $t_{s-d}$ with distance $z$. Therefore, in the case of degeneracy between two $s$-states, the energy interaction is dominated by 
eq.~(\ref{eq:intE1th}) giving rise to the quadratic dependence $I_t \sim E^2$ (see fig.~\ref{Fig:RES3L} a,b). On the other hand, when the $s$ and $d$ states are degenerate, 
the system shows more complicated behavior, as the scaling regime varies with the distance $z$ (see fig.~\ref{Fig:RES3L} c,d). In a far distance regime, 
where the relation $t_{s-d} \gg \frac{t^2_{s-s}}{\Delta_{s-s}}$ holds, the interaction energy is mainly driven by eq.~(\ref{eq:intE1th}). This leads to the quadratic dependence $I_t \sim E^2$. 
At closer distances, whith the hopping between two $s$-levels given by $t_{s-s}$ growing faster than $t_{s-d}$, the interaction between the $s$-states prevails, $t_{s-d} < \frac{t^2_{s-s}}{\Delta_{s-s}}$, and 
the linear dependence between the current and force is established, $I_t \sim E$. Approaching further, the quadratic form  $I_t \sim E^2$ appears again, 
because the hopping between the $s$-like levels  is much larger than the degeneracy parameter $\Delta_{s-s}$, i.e.  $t_{s-s} \gg \Delta_{s-s}$.        
 
%% Figure 23 level model

\section{Conclusions}

In this paper, we have derived a simple theoretical model, which describes the dependence of the tunneling current and the interaction energy on distance. 
Next, we have analyzed the relation between the force and current. Two characteristic scaling regimes have been identified depending on the degeneracy of frontier 
electronic states and strength of the interaction. Our simple model predicts two different scaling regimes between the tunneling current and the interaction energy/force. 
In particular, if we consider the interaction of two non-degenerate and/or delocalized states, the relation $I\sim E$ proposed by Hofer \& Fischer \cite{Hofer03} prevails.
The existence of this regime has been already proven both experimentally \cite{Hembacher05, Bollinger04, Sun05, Ternes11} and theoretically 
using DFT simulations \cite{Hofer03, Ternes11} on metal surfaces. On the other hand, in the case of two strongly localized and degenerate levels, such as dangling bond states on semiconductor surface, the power law  regime $I\sim E^{2}$ prevails.
Indeed, such experimental observation has been already reported on Si surface \cite{Loppacher00}, but the detailed
understanding of its origin has been missing up to now. Moreover, preliminary results of our DFT calculations considering interaction between  a~simple Si tip and adatom 
on the Si(111)-(7x7) surface point out the quadratic relation $I \sim E^{2}$ on the onset of the chemical force. 
The origin of the quadratic relation $I \sim E^{2}$ can be utilized for an~advanced characterization of the electronic states of the tip and sample based on the electron transport between them.

In addition, we discussed the tunneling current dependence on distance using the standard Bardeen approach and the approach where all multiple scattering 
events of electrons passing through the tip-sample junction are taken into account. At far distances both approaches give identical results. However, the Bardeen approach gives unrealistic increase
diverging to infinity at close distances, while the multiple scattering method converges correctly to the ballistic regime. The effect should be taken into account when analysis of 
STM images acquired at low-resistance regime is performed \cite{OndracekPRL11}. 

\section{Acknowledgements}
This work has been supported by the GAAV grant No.\ M100100904  and the GA\v{C}R projects No.\ 204/10/0952, No.\ 202/09/0545 and No.\ 204/11/P578.

% Biblio


\begin{thebibliography}{}
\bibitem{STM}  G. Binnig et al, Phys. Rev. Lett. \emph{49}, 57 (1982).

\bibitem{AFM} G. Binnig et al, Phys. Rev. Lett.  \emph{56}, 930 (1986).

\bibitem{qPLus} F. J. Giessibl, Appl. Phys. Lett. \emph{76}, 1470 (2000).

\bibitem{Datta05} S. Datta  Electronic Transport in Mesoscopic Systems, Cambridge University Press (2005).

\bibitem{Loppacher00} Ch. Loppacher et al Phys. Rev. B \emph{62}, 16944 (2000).

\bibitem{Hembacher05} S. Hembacher et al, Phys. Rev. Lett. \emph{94}, 056101 (2005). 

\bibitem{Bollinger04} G. Rubio-Bollinger et al, Phys. Rev. Lett. \emph{93}, 116803 (2004).

\bibitem{Bollinger01} G. Rubio-Bollinger et al, Phys. Rev. Lett. \emph{87}, 026101 (2001).

\bibitem{Schirmeisen00} A. Schirmeisen et al, New. J. Phys. \emph{2}, 29 (2000). 
 
\bibitem{Sun05} Y.~Sun, H.~Mortensen, S.~Schar, A.~S. Lucier, Y.~Miyahara, P.~Grutter, and
  W.~A. Hofer, Phys. Rev. B \emph{71}, 193407 (2005).

\bibitem{Ternes08} M.~Ternes, C.~P. Lutz, Ch.~F. Hirjibehedin, F.~J. Giessibl, and A.~J. Heinrich.
Science  \emph{319}, 1066 (2008).

\bibitem{Konig09} T.~Konig, G.H.~Simon, H.-P. Rust and M. Heyde App. Lett. \emph{95}, 083116 (2009).

\bibitem{Ternes11} M.~Ternes, C~. Gonz\'{a}lez, Ch.~P. Lutz, P~. Hapala, F.~J. Giessibl, P.~Jel\'{i}nek and A.~J. Heinrich,
Phys. Rev. Lett.  \emph{106},  176101 (2011).

\bibitem{Sawada09a} D.~Sawada, Y.~Sugimoto, K.~Morita, M.~Abe, and S.~Morita,
Appl. Phys. Lett., \emph{94}, 173117 (2009).
 
\bibitem{Cui01}  X.D.~Cui, A.~Primak, X.~Zarate, J.~Tomfohr, O.F.~Sankey, A.L.~Moore, T.A.~Moore, D.~Gust, 
G.~Harris and S.M~Lindsay, Science, \emph{294}, 571 (2001).
      
\bibitem{VenkataramanNature06} L.~Venkataraman, J.E.~Klare, C.~Nuckolls, M.S.~Hybertsen and M.L.~Steigerwald
Nature, \emph{442}, 904 (2006).

\bibitem{Tautz_PRB11} N.~Fournier, C.~Wagner, C.~Weiss, R.~Temirov and F.S.~Tautz
Phys. Rev. B,  \emph{84}, 035435 (2011).

\bibitem{Nitzan03} A.~Nitzan, M.A.~Ratner, Science \emph{300}, 1384 (2003).

\bibitem{Schwab05} K.C.~Schwab, M.L.~Roukes, Phys. Today \emph{58}, 36 (2005).
 
\bibitem{Bardeen60} J.~Bardeen Phys. Rev. Lett. \emph{6}, 57 (1960). 
 
\bibitem{Lowdin63}P.O. L\"{o}wdin, J.Mol. Spectroscopy \emph{10},
12 (1963).

\bibitem{Goldberg89} E.C. Goldberg et al, Phys. Rev. B \emph{39},
5684 (1989).

\bibitem{Vidal91}F.J. Garcia-Vidal et al, Phys. Rev. B \emph{44},
11412 (1991).

\bibitem{Ortega94} J.~Ortega et al, Phys. Rev. B \emph{50},
10516 (1994).

\bibitem{MingoPRB96} N.~Mingo et al, Phys. Rev. B \emph{54},
2225 (1996).

\bibitem{Blanco06}J.M.~Blanco, F.~Flores and R.~Perez, Prog. Surf. Sci. \emph{81},
403 (2006).

\bibitem{Dappe06}Y.~Dappe et al, Phys. Rev. B \emph{74}, 205434
(2006).

\bibitem{Basanta05}M.A.~Basanta et al, Europhys. Lett. \emph{70}, 355
(2005).

\bibitem{Hofer03}W.~Hofer and A.J.~Fisher Phys. Rev. Lett. \emph{91}, 036803 (2003).

\bibitem{Neel07} N.~N\'{e}el, J. Kr\"{o}ger, L. Limot, K. Palatos, W. Hofer and R. Berndt,  Phys. Rev. Lett. \emph{98}, 016801 (2007).

\bibitem{Chen91}C.J.~Chen J.Phys.: Cond. Matt. \emph{3}, 1227
(1991).

\bibitem{Chen93} C.J.~Chen Introduction to Scanning Tunneling Microscopy, Oxford University Press (1993).

\bibitem{Harrison99} W.A.~Harrison Elementary Electronic Structure, World Scientific Publishing (1999).

\bibitem{Chen05} C.J.~Chen Nanotechnology \emph{16}, S27 (2005).

\bibitem{Chen91a} C.J.~Chen, R.J.~Hamers, J. Vac. Sci. Technol. B \emph{9}, 503 (1991).

\bibitem{Oppenheimer29} J.R.~Oppenheimer Phys. Rev. \emph{13}, 66 (1928).

\bibitem{Sugimoto_PRB06} Y. Sugimoto, P. Pou,  O. Custance, P. Jel\'{i}nek,  S. Morita,  R. P\'{e}rez, M. Abe,  Phys. Rev. B \emph{73}, 205329 (2006). 

\bibitem{Sugimoto-Nature07} Y. Sugimoto, P. Pou, M. Abe, P. Jel\'{i}nek, R. P\'{e}rez, S. Morita, O. Custance  , Nature \emph{446}, 64 (2007).

\bibitem{Pou-Nanotechnology09}  P. Pou, S.A. Ghasemi, P. Jel\'{i}nek, T. Lenosky, S. Goedecker, R. P\'{e}rez  
Nanotechnology \emph{20},  264015 (2009).

\bibitem{Landauer70} R.~Landauer, Philos. Mag. \emph{21}, 863 (1970).

\bibitem{JelinekSS04} P. Jel\'{i}nek, R. P\'{e}rez, J. Ortega and F. Flores, Surf. Sci. \emph{566-568}, 13 (2004). 

\bibitem{TransSiesta} M.~Brandbyge, J.L.~Mozos, P.~Ordejon, J.~Taylor and K.~Stokbro, Phys. Rev. B \emph{65}, 165401 (2002).

\bibitem{Smeagol} A.R.~Rocha, V.~Garcia-Suarez, S.W.~Bailey, C.J.~Lambert, J.~Ferrer and S.~Sanvito, Phys. Rev. B \emph{73}, 085414 (2006).

\bibitem{Blanco_PRB04} J.M. Blanco, C. Gonz\'{a}lez, P. Jel\'{i}nek, J. Ortega, F. Flores and R. P\'{e}rez Phys. Rev. B \emph{70}, 085405 (2004). 

\bibitem{Jelinek_PRL08} P.Jel\'{i}nek, M. \v{S}vec, P. Pou, R. P\'{e}rez and  V. Ch\'{a}b Phys. Rev. Lett. \emph{101}, 176101 (2008).

\bibitem{Lang88} N.~D. Lang. Phys. Rev. B, \emph{37}, 10395 (1988).

\bibitem{ChoPRL07} Y.~Cho and R.~Hirose, Phys. Rev. Lett. \emph{99}, 186101 (2007). 

\bibitem{BevanPRB08} K.H.~Bevan, D.~Kienle, H.~Guo and S.~Datta, Phys. Rev. B \emph{78}, 035303 (2008). 

\bibitem{Sadewasser09} S.~Sadewasser, P.~Jel\'{i}nek, C.K.~Fang, O.~Custance, Y.~Yamada, Y.~Sugimoto, M.~Abe, and S.~Morita,
Phys. Rev. Lett. \emph{103}, 266103  (2009).

\bibitem{OndracekPRL11} M.~Ondr\'{a}\v{c}ek, P.~Pou, C.~Gonzalez, P.~Jel\'{i}nek and R.~Perez, Phys. Rev. Lett. \emph{106}, 176101 (2011).
\end{thebibliography}
\end{document}